\documentclass[prd,aps,reprint,amsmath,amssymb,eshowkeys,showkeys,showpacs,nofootinbib]{revtex4-1}
\pdfoutput=1
\usepackage{verbatim,graphics,graphicx,color,slashed,textcomp,bbm,mathdots}
\usepackage[normalem]{ulem} 
\usepackage[colorlinks=true,linkcolor=red,urlcolor=blue,citecolor=blue]{hyperref}

\graphicspath{{figures/}}

\usepackage[utf8]{inputenc}

\begin{document}

\title{Probing Dark Matter Spikes via Gravitational Waves \\ of Extreme-Mass-Ratio Inspirals }
\author{Gen-Liang Li$^{a,c}$}
\author{Yong Tang$^{a,b,c,d}$}\email{Corresponding author, tangy@ucas.ac.cn}
\author{Yue-Liang Wu$^{a,b,c,e}$}
\affiliation{\begin{footnotesize}
		${}^a$University of Chinese Academy of Sciences (UCAS), Beijing 100049, China\\
		${}^b$School of Fundamental Physics and Mathematical Sciences, \\
		Hangzhou Institute for Advanced Study, UCAS, Hangzhou 310024, China \\
		${}^c$International Center for Theoretical Physics Asia-Pacific, Beijing/Hangzhou, China \\
		${}^d$National Astronomical Observatories, Chinese Academy of Sciences, Beijing 100101, China\\
		${}^e$Institute of Theoretical Physics, Chinese Academy of Sciences, Beijing 100190, China
\end{footnotesize}}

\begin{abstract}
	The exact properties of dark matter remain largely unknown despite the accumulating evidence. If dark matter is composed of weakly interacting massive particles, it would be accreted by the black hole in the galactic center and form a dense, cuspy spike. Dynamical friction from this spike may have observable effects in a binary system. We consider extreme-mass-ratio inspiral (EMRI) binaries comprising massive black holes harbored in dark matter spikes and stellar mass objects in elliptic orbits. We find that the gravitational-wave waveforms in the frequency domain can be substantially modified. In particular, we show that dark matter can suppress the characteristic strain of a gravitational wave at low frequency but enhance it at a higher domain. These effects are more dramatic as the dark matter density increases. The results indicate that the signal-to-noise ratio of EMRIs can be strongly reduced near $10^{-3}\sim 0.3$~Hz but enhanced near $1.0$~Hz with a higher sensitivity, which can be probed via the future space-borne gravitational-wave (GW) detectors, LISA and TAIJI. The findings will have important impacts on the detection and parameter inference of EMRIs.
	
\end{abstract}


\maketitle

\section{Introduction}
Dark matter (DM) has been a challenging problem in modern astronomy, cosmology, and physics. Despite convincing evidence from galactic to cosmological observations~\cite{Corbelli:1999af,Clowe:2006eq,Planck:2018vyg}, DM's particle nature is still speculative. As one of the popular candidates for DM, the thermal weakly interacting massive particle (WIMP) is very attractive in terms of detection directly and indirectly~\cite{Bertone:2004pz,Ibarra:2013cra,Schumann:2019eaa}. WIMPs form a universal density profile in galactic halos~\cite{Navarro_1997}, which can be very cuspy near the galactic center and even constitute a spike near the central massive black hole (BH)~\cite{SilK_1999}, affecting the cosmic ray searches of DM~\cite{Lacroix:2013qka}. Although major merger events of host galaxies~\cite{Merritt:2002vj} and scattering of dark matter particles by surrounding stars~\cite{Merritt:2003qk,Bertone:2005hw} can possibly weaken and even destroy the spikes, many DM minispikes might be surrounding the extramassive black holes, with a mass range between $10^2$ and $10^6 M_{\odot}$ surviving if the BH never experienced any major mergers~\cite{Zhao:2005zr}.

Recently, the possibility of probing DM with gravitational-wave (GW) experiments has been proposed. In~\cite{SilK_2013, Macedo:2013qea}, dynamical friction from DM spikes was shown to change the phase of a GW in extreme-mass-ratio inspiral (EMRI) binaries with circular orbits and reduce the signal significance if waveform templates without DM are used for matched filtering. However, others~\cite{Barausse_2014} state that the corrections of DM halos to the GW signal or environmental effects are negligible. Later, studies with more effects~\cite{Silk_2015, Yue:2017iwc, Yue:2018vtk, Traykova:2021dua, Chung:2021roh} continue and support the claim that the phases of GWs can be modified to have impacts on future space-borne GW experiments. Extensions to different DM candidates~\cite{Hannuksela:2018izj, Hannuksela:2019vip, Edwards:2019tzf, Ferreira:2017pth} and discussions of back-reaction~\cite{Kavanagh:2020cfn, Coogan:2021uqv, Traykova:2021dua} have also been explored. However, it remains unclear how the significance of a signal is modified if the waveform templates with DM are used for matched filtering.

The above investigations only focus on circular orbits and the phase of a GW. However, EMRIs are more likely to form from elliptic orbits and be detected by future LISA~\cite{LISA:2017pwj} and TAIJI~\cite{Hu:2017mde, Ruan:2020smc, Wu:2021}. A more realistic approach would be to estimate how GW signals from elliptic orbits can be affected by dynamical friction. Refs.~\cite{Yue:2019ozq, Tang_2021} show that a DM spike can enhance the eccentricity and amplitude of a GW waveform in the {\it time} domain. Therefore, enhancement of a GW signal on LISA and TAIJI would be expected, which, however, seems to contradict the above conclusions because circular orbits are simply special cases with zero eccentricity.

In this paper, we study the impact of dynamical friction on the GW's characteristic strain and signal-to-noise ratio ($S/N$) by numerically solving the coupled dynamical equations for EMRIs in elliptic orbits and then analyze the effects in the {\it frequency} domain. We first show that dynamic friction can suppress the amplitude of a gravitational wave at a low frequency but enhance it at a higher domain, and the $S/N$ of EMRIs can be strongly reduced near $10^{-3}\sim 0.3$~Hz but enhanced near $1.0$~Hz with a higher sensitivity, which can be probed via the future space-borne GW detectors, LISA and TAIJI. We suggest extending the sensitivity of detectors at a higher frequency near $1.0$~Hz for future space-borne GW detectors. These effects will have important impacts on the detection and parameter inference of EMRIs and provide a potential probe of DM.

This paper is organized as follows. In section~\ref{sec:DM} we describe how to determine the DM density profile near a BH, consider dynamical friction when evolving EMRI systems, and expand the characteristic strain in harmonics. Later, in Section~\ref{sec:GW}, we investigate how DM can affect the GW spectra in the sensitive frequency domain of future detectors. Then, in Section~\ref{sec:sn}, we illustrate how the effects can modify the $S/N$ of EMRIs and influence the detection and parameter inference of this system. Considering the uncertainty that DM particles can have different velocities, we also compare the possible influence of velocity dispersion on the characteristic spectra and the S/N of GWs in Section~\ref{sec:VD}. Finally, we give our conclusion.

\section{DM Density and GW Emission}\label{sec:DM}

\begin{figure}[t]
	\centering
	\includegraphics[width=8cm]{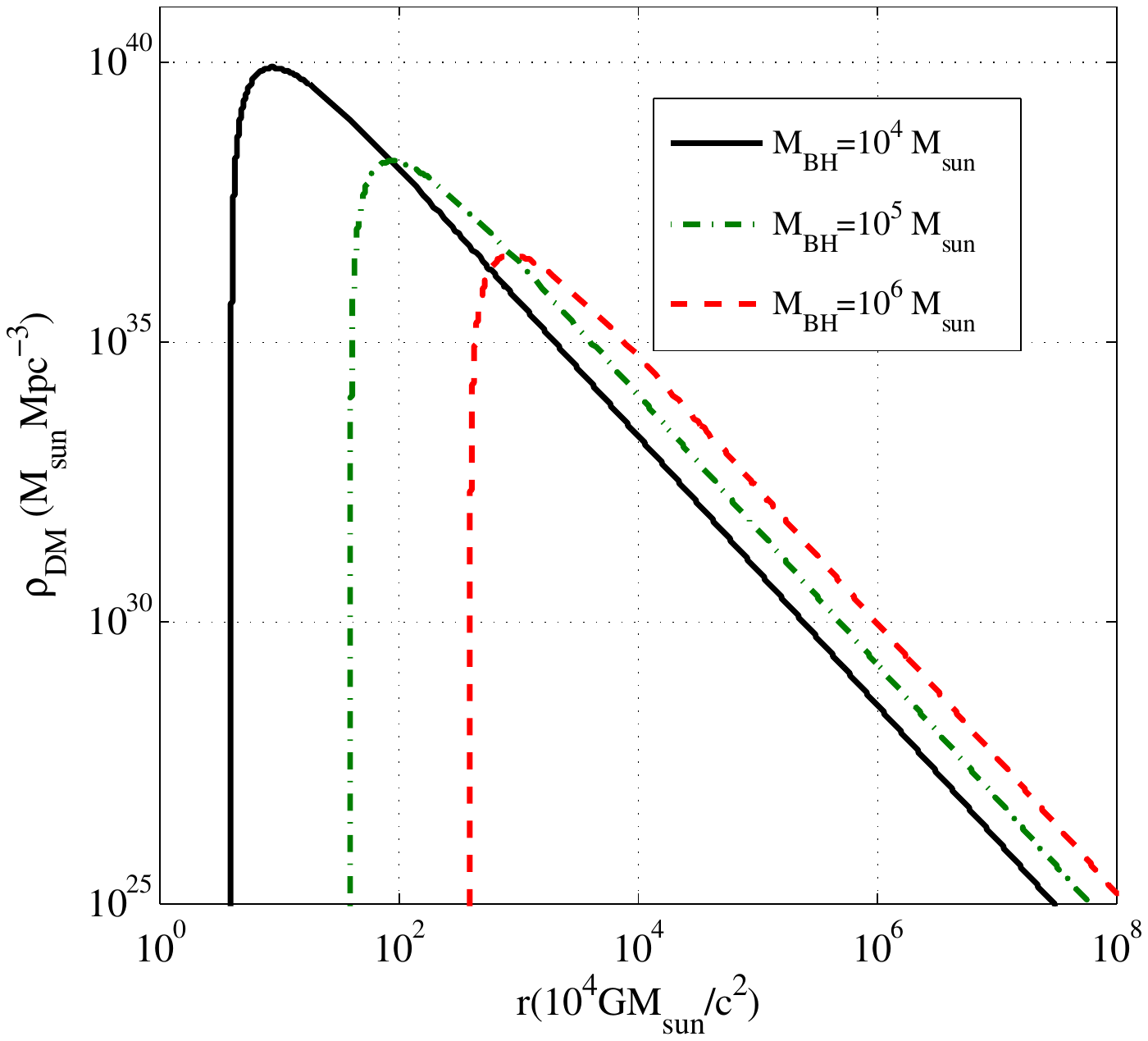}   
	\caption{\label{fig:diff_M_density} DM density profile $\rho_{_\textrm{DM}}$ with $\alpha=2.4$ as functions of the distance to the galaxy center with different black hole masses $M_{\textrm{BH}}$. The distance $r$ is normalized to $10^4GM_{\textrm{sun}}/c^2$. }
\end{figure}

We consider a massive BH dressed initially by a DM profile $\rho_{i}\propto \rho_0 (r_0/r)^\gamma$ and growing adiabatically. A cuspy profile would result from the gravitational concentration and is usually referred to as a DM spike~\cite{SilK_1999}. We consider the adiabatic growth of the Schwarzchild BH with relativistic effects taken into account~\cite{sadeghian2013dark} and describe the DM density distribution near the BH with mass $M_{\textrm{BH}}$ as
\begin{equation}\label{eq:density}
	\rho_{_\textrm{DM}}(r)=\rho_{sp}\left(1-\frac {2R_s}{r}\right)^3\left(\frac{r_{sp}}{r}\right)^{\alpha},
\end{equation}
where the derived parameter $\rho_{sp}=\rho_0(\frac{r_0}{r_{sp}})^{\gamma}$, $r_{sp}=\beta_{\gamma}r_0\left(\frac{M_{\textrm{BH}}}{\rho_0 r_0^3}\right)^{{1}/{(3-\gamma)}}$, $\alpha=\frac{9-2\gamma}{4-\gamma}$, $R_s=2GM_{\textrm{BH}}/c^2$ is the Schwarzschild radius, and $G$ is Newton's constant. We consider the power index $\gamma \in [0,2]$ for the initial DM profile, which gives $\alpha \in [2.25,2.5]$. We determine the initial ${\rho_0}$ and ${r_0}$ through the mass-velocity-dispersion relationship following~\cite{nishikawa2019primordial}, which correlates $M_\textrm{BH}$ and its host halo mass. We then calculate the corresponding derived parameters for ${\rho_{_\textrm{DM}}}$. In Fig.~\ref{fig:diff_M_density}, we plot the DM density near the BH with $\alpha=2.4$ in three cases with different BH masses. These curves show explicitly that the density increases toward the galactic center but suddenly drops near $2R_s$ because of the factor $\left(1- {2R_s}/{r}\right)^3$, which arises because of the absence of a stable orbit near the BH horizon. This figure also demonstrates that DM density near the horizon is larger for a smaller BH, which would have distinguishable effects on the GW, as we shall show later.

The binary system of a neutron star or a stellar BH with $m_2$ orbiting a massive BH with $m_1$ surrounded by a DM spike is generally elliptic, with eccentricity $e$ and orbital frequency $f_{\textrm{orb}}$ (in the rest-frame of the binary). The system would lose energy $E$ and angular momentum $L$ due to GW emission and dynamical friction (DF) from DM,
\begin{equation}
	\frac{dE}{dt}= \langle \frac {dE}{dt }\rangle _{_\textrm{GW}} +  \langle \frac {dE}{dt} \rangle _{_\textrm{DF}},\;
	\frac{dL}{dt}= \langle \frac {dL}{dt }\rangle _{_\textrm{GW}}  +  \langle \frac {dL}{dt} \rangle _{_\textrm{DF}},
\end{equation}
where $ \langle \cdot \rangle $ denotes the average loss of energy and angular momentum in a period due to GW radiation and DF, respectively. For the GW emission, we use the post-Newtonian formalism and keep the leading term~\cite{Peters_1963} for our discussions. The contributions from DF involve complicated integration over one period; therefore, they are evaluated numerically. The relevant formulas are summarized in the appendix, Eqs.~\ref{eq:en}-\ref{eq:e}.
Here and hereafter, we neglect other subdominant effects, such as the next-to-leading order post-Newtonian contribution, dynamical hardening~\cite{G_rkan_2006}, and accretion~\cite{Yue:2019ozq}. As we shall show, the effect of DF can be larger than the dominant GW contribution, including those of other subdominant terms, which would not change our results qualitatively.

For a direct comparison with the sensitivities of space-borne GW detectors, such as TAIJI and LISA, we shall generate the GW spectra in the frequency domain. We implement the formalism by ~\cite{Finn:2000sy, BC04} and calculate the GWs by summing the higher harmonics due to the elliptic orbit. The frequency-domain GW spectra for $n$-order harmonic waves and frequency $f_{n}=nf_{\textrm{orb}}$ has the power of harmonics~\cite{Peters_1963}:
\begin{eqnarray}\label{eq:dEn}
	\frac {dE_n}{dt} =\frac {G^{{7}/{3}}}{5c^5}(2 \pi M_c f_{\textrm{orb}})^{{10}/{3}}g_n(e),
\end{eqnarray}
where $M_c=(m_1m_2)^{3/5}/(m_1+m_2)^{1/5}$ is the chirp mass, and $g_n$ is a Bessel function of the first kind (see Eq.~\ref{eq:gn} in the appendix). The characteristic strain of each harmonic is given by
\begin{eqnarray}\label{eq:Hcn}
	h_{c,n}=\frac{1}{\pi d}\sqrt{\frac{2G \dot{E_n}}{c^3\dot{f_n}}}.
\end{eqnarray}
Here, the dot means the time derivative, $\dot{f_n}=n\dot{f}_{\textrm{orb}}$ is the evolution of the $n$-th harmonic, and $d$ is the distance from the source binary to the solar system. For sources at redshift $z$, the rest-frame orbital frequency $f_{\textrm{orb}}$ and observed frequency $f$ satisfies $f_{\textrm{orb}}=f(1+z)$ and $f_n=nf_{\textrm{orb}}$. We then obtain the characteristic strain by summing over all harmonics
$
h^2_{c}(f)= \sum_{n=1}^{\infty}h^2_{c,n}.
$

\begin{figure*}[tb]
	\centering
	\begin{minipage}[t]{\columnwidth}
		\includegraphics[width=1.05\textwidth,height=\textwidth]{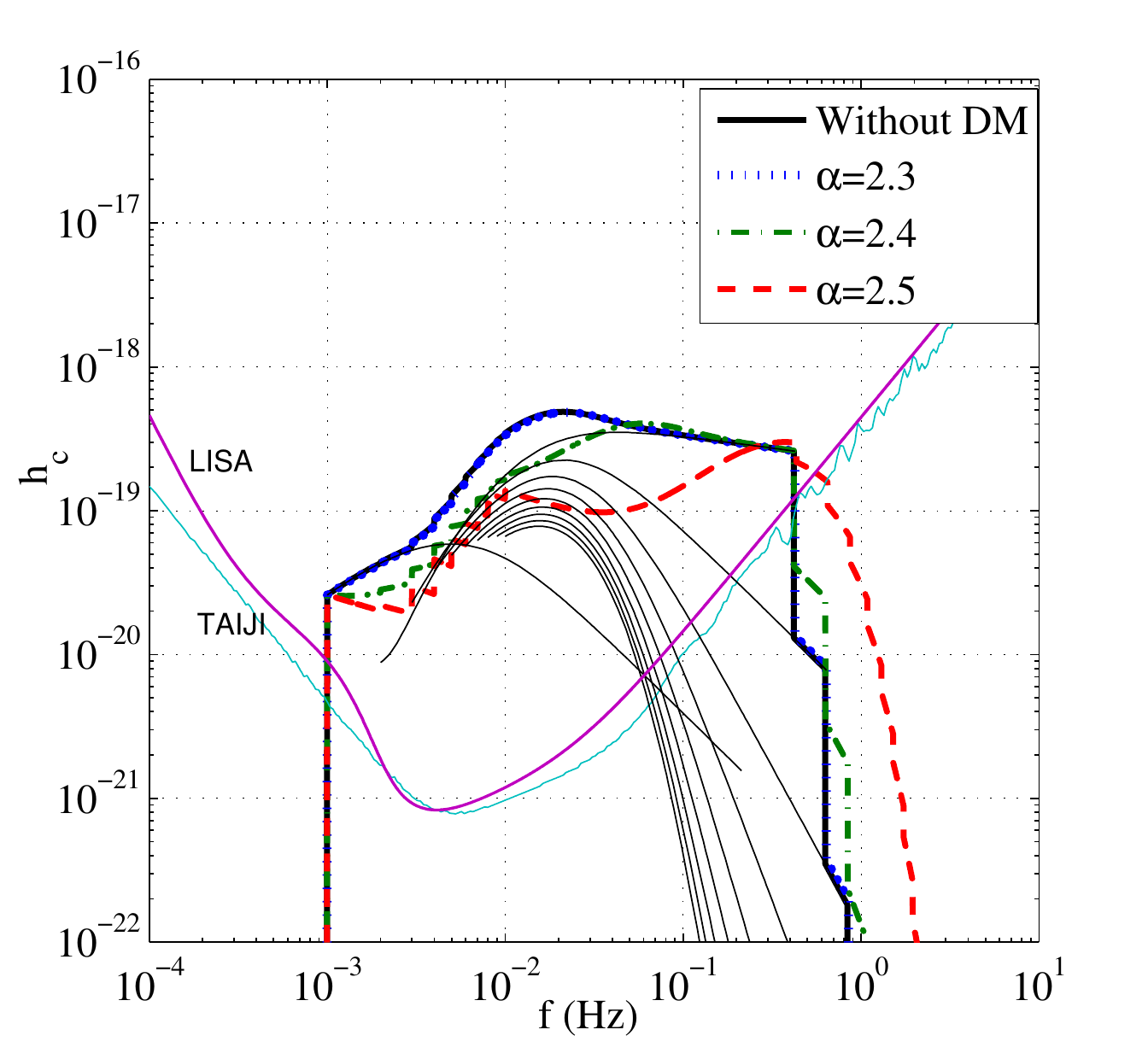}
	\end{minipage}
	\begin{minipage}[t]{\columnwidth}
		\includegraphics[width=1.05\textwidth,height=0.95\textwidth]{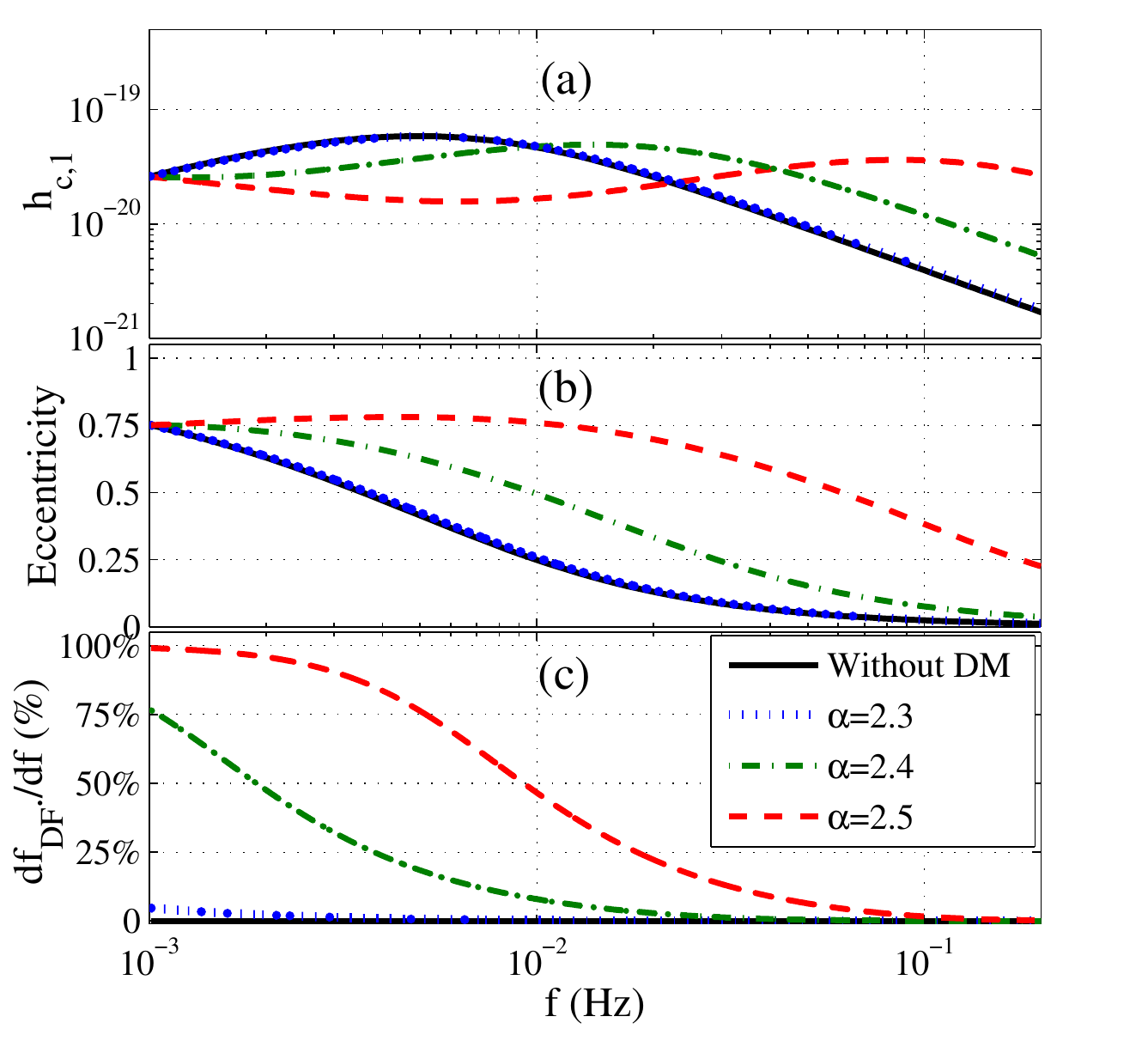}
	\end{minipage}  
	\caption{\label{fig:harmonics} (Left) GW characteristic strain and effects from DM's DF for an EMRI at $z=0.01$. For illustration, we choose $m_1=10^{4}M_\odot$, $m_2=10 M_\odot$, the initial orbital frequency at $f_{\textrm{ini}}=10^{-3}$~Hz, and eccentricity $e_0=0.75$. The solid black thin curves indicate the different harmonics, which are summed to give the solid black thick curve without DM. When DM is considered, the corresponding strains are plotted as dotted blue ($\alpha=2.3$), dot-dashed green ($\alpha=2.4$), and dashed red ($\alpha=2.5$) curves, respectively. Sensitivities from LISA and TAIJI are shown as the purple and light blue curves. (Right) Evolutions of harmonic $h_{c,1}$, eccentricity, and the relative contribution of DF as functions of orbital frequency. }
\end{figure*}

\section{Effects on the GW Spectra}\label{sec:GW}

Now, we can quantify the effects on the GWs from DM spikes. To make our discussions more concrete, we first illustrate a binary system with an MBH of $m_1=10^4M_{\odot}$ and an inspiring BH with $m_2=10M_{\odot}$ ($M_\odot=M_\textrm{sun}$ is the Sun's mass). With initial inputs of $f_{\textrm{ini}}$ and $e_0$, we numerically solve the coupled differential--integral equations for the orbital frequency $f_\textrm{orb}$ and eccentricity $e$, from which we can obtain the values of the characteristic strains for each harmonic.

We assume a four-year observational time for a space GW detector. If the EMRI binary does not coalesce within 4 years, we truncate at a maximum frequency of $f_{\textrm{max}}=f(t=4~\textrm{years})$. Otherwise, we can terminate the evolution at a maximum frequency of $f_{\textrm{max}}=c^3/(2\pi x^{3/2}GM_{\textrm{BH}})$, with the factor $x\simeq 6$, as we are considering the nonspinning Schwarzschild BH~\cite{Bonetti:2020jku}.

In the left panel of Fig.~\ref{fig:harmonics}, we show the typical characteristic strain of a GW from EMRIs with/without DM spikes. We choose a binary source placed at a redshift of $z=0.01$ with the initial values $f_{\textrm{ini}}=10^{-3}$~Hz and $e_0=0.75$. Different harmonic contributions up to $n=10$ are plotted as solid black {\textit{thin}} curves, which are summed to obtain the total characteristic strain $h_c$ as the solid black \textit{thick} curve. Each harmonic can be identified as a thin black curve with a different starting frequency $f_n=nf_\textrm{ini}$. Unlike the cases with circular orbits where only the quadruple mode $n=2$ contributes, GWs from elliptic orbits have a wide band in the frequency domain, and many harmonics can be comparable in terms of strain at some frequency range. The relative importance of each harmonic also depends on the eccentricity $e$, as we shall show shortly.

Next, we investigate how DM spikes can modify the GW spectra. In the same plot in Fig.~\ref{fig:harmonics} we consider the DM density profile with three power indices, $\alpha=2.3, 2.4$, and $2.5$, and show the total GW spectra as dotted blue, dot-dashed green, and dashed red curves, respectively. The initial conditions are identical, $f_{\textrm{ini}}=10^{-3}$~Hz and $e_0=0.75$, but the spectra can be substantially different in some cases. When $\alpha=2.3$ or $\gamma=2/3$, the dotted blue curve is almost identical to the black one without DM. However, as we increase the index $\alpha$ or $\gamma$, we see the sizable deviations that can distinguish DM profiles. Two general features appear in the green and red curves in comparison with the black one and reflect two important effects on GW emission from DM's DF. The first feature is that the GW spectrum is broader with DM in action, and the second feature is that the strain is generally suppressed at lower frequencies but enhanced at a higher domain.

The above two effects have a common physical reason: The DF tends to increase eccentricity or slow down the decline rate, while the GW emission alone decreases $e$ monotonously and circularizes the orbits. To verify this explanation, in the right panel of Fig.~\ref{fig:harmonics} we plot comparably the corresponding $n=1$ harmonic $h_{c,1}$, eccentricity $e$, and the relative contribution of dynamical friction to the frequency evolution $df_{_{\textrm{DF}}}/df$. The top figure clearly shows the effects on the characteristic strain $h_{c,1}$, namely, suppression at low frequency and enhancement at a higher range. Similar patterns are also applied to other harmonics. These behaviors reflect the modification of the eccentricity evolution, as shown in the middle figure. As we can observe, when the index $\alpha$ increases, $e$ decreases more slowly than in the case without DM and can even increase for large $\alpha$. $e\rightarrow 0$ without DM as the system approaches the plunge due to the circularization of GW emission, whereas $e$ with DM can be sizable before the plunge. The bottom figure indicates that in the early phase of evolution, DF dominates the back-reaction from GW emission. As the system evolves, GW gradually dominates and starts to decrease $e$ and circularize the orbit. The relative importance of DF and GW emission depends on the DM density profile and the distance between the binary components or the orbital frequency.

Observing that the strain in the frequency domain depends on the Fourier transform is also helpful for understanding the above effects in the time domain:
\begin{eqnarray}
	h(f) = \int_0^{t_c}{h(t)e^{-i2\pi ft} dt},
\end{eqnarray}
where $h(t)$ is the time-domain GW, and $t_c$ is the coalescence time. If we have an EMRI in a circular orbit, the spectrum will almost be monochromatic without orbital evolution. The actual strain size depends on how much time the system has stayed at the frequency $f_{\textrm{orb}}$. If the orbit is elliptic, then different harmonics can be obtained by decomposing the orbit as the superposition of monochromatic components. The strain for each harmonic depends on how much time the corresponding component can last. When we include the DM's DF, the binary system accelerates, and the evolution time scale is shortened. Then, the low-frequency components will be relatively suppressed, and the high-frequency ones will be enhanced.

\section{Signal-to-Noise Ratio}\label{sec:sn}

\begin{figure*}[t]
	\centering
	\begin{minipage}[t]{\columnwidth}
		\includegraphics[width=\textwidth,height=0.95\textwidth]{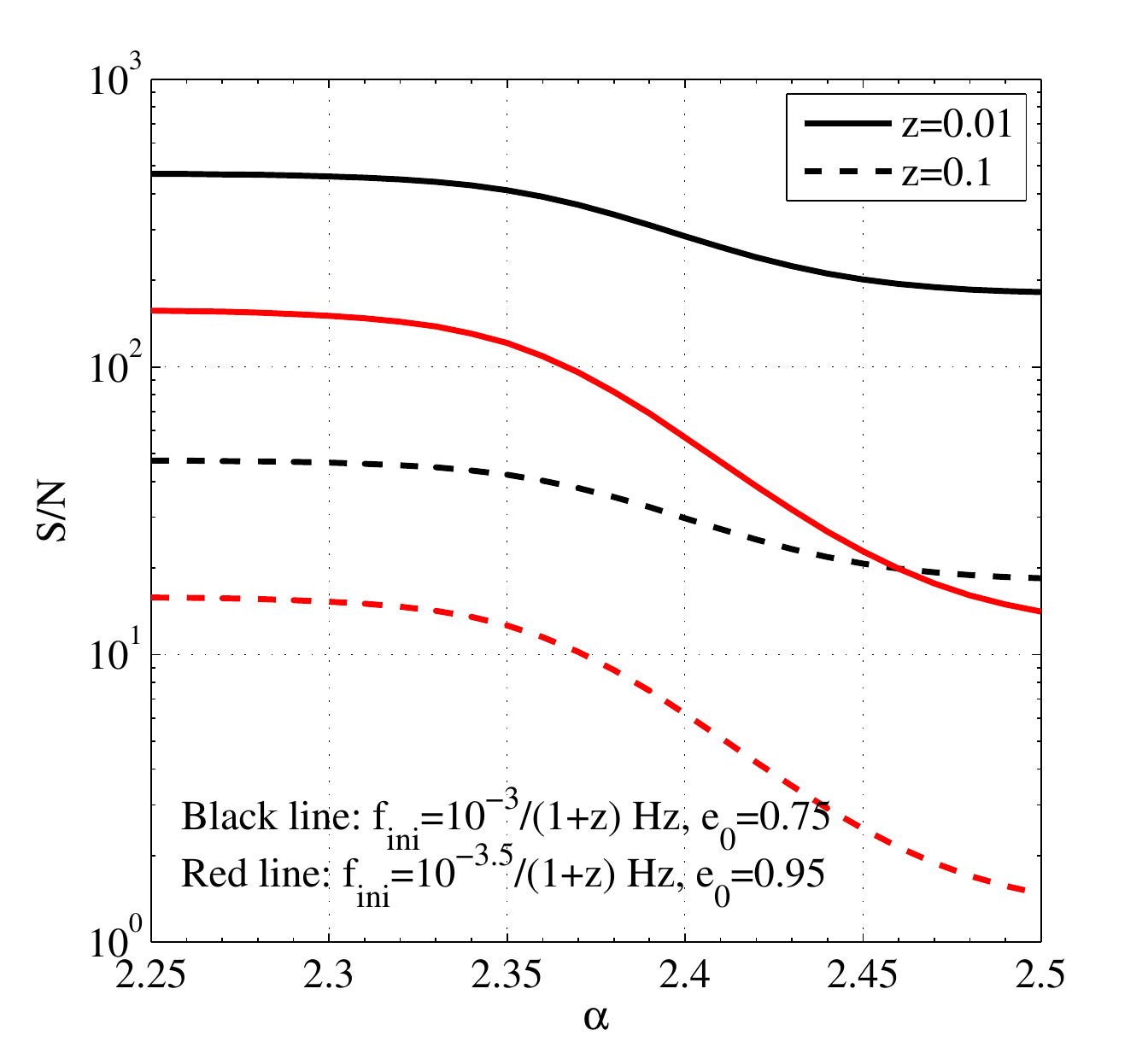}
	\end{minipage}
	\begin{minipage}[t]{\columnwidth}
		\includegraphics[width=1.1\textwidth,height=0.95\textwidth]{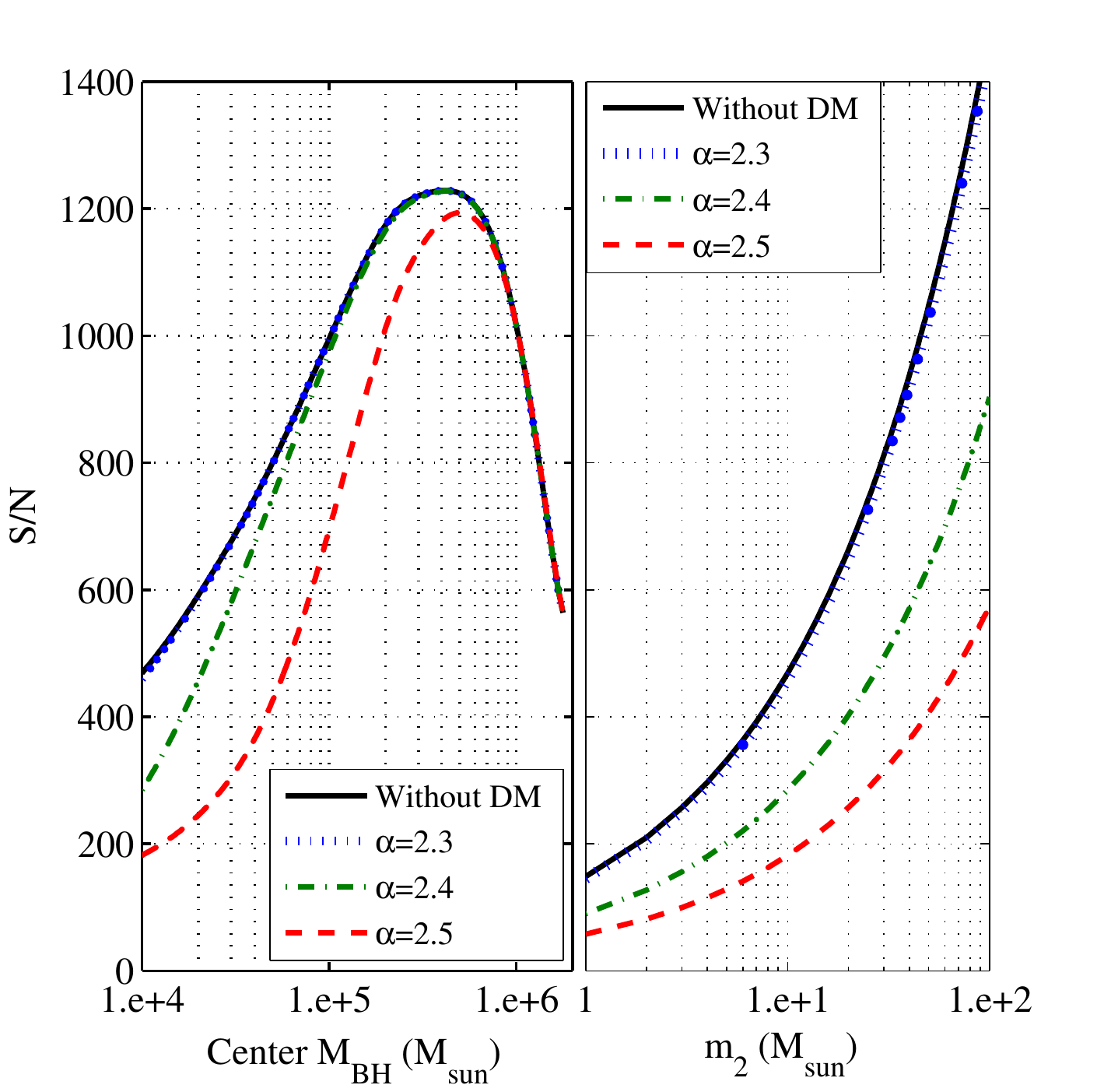}
	\end{minipage}  
	\caption{\label{fig:SNR}(Left) $S/N$ against the DM density index $\alpha$. All curves have $m_1=10^{4}M_\odot$ and $m_2=10 M_\odot$. The black lines (solid for $z=0.01$ and dashed for $z=0.1$) are obtained with $f_{\textrm{ini}}=10^{-3}/(1+z)$~Hz and $e_0=0.75$, while the red ones (solid for $z=0.01$ and dashed for $z=0.1$) are obtained with $f_{\textrm{ini}}=10^{-3.5}/(1+z)$~Hz and $e_0=0.95$. (Right) The $S/N$ for the EMRI binary at $z=0.01$ as a function of the mass of central BH and secondary objects (we fix the inspiring body mass as $m_2 = 10 M_{\odot}$ (the first one) and the central BH mass as $M_{\textrm{BH}} = 10^4 M_{\odot}$ (the second one)), with initial conditions $f_{\textrm{ini}}\simeq 10^{-3}$~Hz, $e_0=0.75$, and $z=0.01$. The solid black curve describes the case without DM; the dotted blue one, $\alpha=2.3$; the dot-dash green one, $\alpha=2.4$; and the dashed red one, $\alpha=2.5$.}
\end{figure*}

Now, we discuss how the above effects can change the signal-to-noise ratio of EMRIs at space-borne GW detectors. After obtaining the characteristic strain, the resulting $S/N$ is then calculated using
\begin{eqnarray}\label{eq:SNR}
	(S/N)^2= \sum_{n=1}^{\infty}\int{\frac {h^2_{c}}{f^2_n S_{n}(f_n) }d f_n}.
\end{eqnarray}
Here, $S_n(f_n)$ is the spectral density of TAIJI~\cite{luo2021taiji} or LISA~\cite{LISA:2017pwj}. The sensitivity curves from TAIJI and LISA are indicated in Fig.~\ref{fig:harmonics} as light blue and solid purple curves, respectively. Because they are qualitatively at the same level, we shall mainly refer to TAIJI for the later discussions.

In the left panel of Fig.~\ref{fig:SNR}, we show how the values of $S/N$ change with the DM density profile index $\alpha$ in Eq.~\ref{eq:density}. We choose four illustrative examples with two initial orbital frequencies and eccentricities at two redshifts $z=0.01$ (solid) and $0.1$ (dashed), $f_{\textrm{ini}}=10^{-3}$~Hz and $e_0=0.75$ (black curves), and $f_{\textrm{ini}}=10^{-3.5}$~Hz and $e_0=0.95$ (red curves). These curves suggest a general pattern in which increasing the profile index $\alpha$ leads to suppression of the $S/N$, although the details can differ. For instance, for the top black curve, as $\alpha$ approaches $2.5$, $S/N$ changes more slowly. This result is due to the lower sensitivity of the space GW detector at high frequency, as shown in Fig.~\ref{fig:harmonics}. When the enhanced strain at high frequency is below the sensitivity curve, it will not substantially contribute to $S/N$. On the other hand, for the red curves with a lower starting frequency, the reduction of $S/N$ can be more dramatic because their dominant strains are mainly in the frequency domain of sensitivity.

Reducing $S/N$ can have substantial effects on EMRI detection and parameter inference. Our results show that $S/N$ may be reduced by up to one order of magnitude. If an EMRI is on the edge of detection threshold $S/N\sim 8$, it may be completely lost if the DF of the DM spike is considered, although the precise value of $S/N$ depends on the DM density index $\alpha$. As a result, the detectable cosmic volume will also be modified. In addition, because the precision of parameter inference is proportional to $1/(S/N)$, it is straightforward to see that determining the parameters of the binary source will also be affected.

Finally, we discuss the effects on $S/N$ change of different masses of central BHs and inspiral objects with a fixed initial frequency and eccentricity. In the right panel of Fig.~\ref{fig:SNR}, we show how DF from DM spikes affects $S/N$ for EMRI systems with different mass ratios. First, we fix the mass of the secondary object $m_2 = 10 M_{\odot}$ and change the mass ratio of the binary system by changing the central BH mass with the initial parameter $f_{\textrm{ini}}=10^{-3}/(1+z)$~Hz, $e_0=0.75$, and $z=0.01$. In general, reducing $S/N$ is less significant in a more massive BH system. This result is due to the larger DM density of a smaller BH near a BH horizon because of the inverse power law in Eq.~\ref{eq:density}. Another feature we notice is that $S/N$ decreases as the BH mass exceeds $\sim 6\times 10^5 M_{\textrm{sun}}$, which is due to the time evolution of the EMRI system. Because the maximum frequency before the plunge $f_{\textrm{max}}\simeq c^3/(2\pi 6^{3/2}GM_{\textrm{BH}})$ is inversely proportional to the mass, more massive systems have a shorter inspiral time before reaching the plunge phase. Therefore, the signal strain will be reduced for EMRIs with larger central BHs. Second, we also adjust the parameter $m_2$ and fix the central black hole mass $M_{\textrm{BH}}=1\times 10^4 M_{\odot}$ with the same initial conditions. One can see S/N increase with $m_2$ or reduced mass $\mu = m_1m_2/(m_1+m_2)\simeq m_2$. This result can be easily explained as follows. According to Eq.~\ref{eq:dEn}, Eq.~\ref{eq:Hcn}, and Eq.~\ref{eq:f}, we can obtain the relationship between GW strains and reduced mass, which is $h_{c,n} \propto \mu ^ {1/2}$. Substituting this relationship into Eq.~\ref{eq:SNR}, we can obtain $S/N \propto \mu ^ {1/2}$. Notably, when the mass ratio is large enough, the minispikes are more likely to be destroyed by the process of the merger of host galaxies~\cite{Merritt:2002vj}. Therefore, extrapolation to a higher mass ratio of $m_2/m_1 \gtrsim 0.01$ should be done with caution.

\section{Effects of Velocity Dispersion}\label{sec:VD}

\begin{figure*}[t]
	\centering
	\begin{minipage}[t]{\columnwidth}
		\includegraphics[width=\textwidth,height=0.95\textwidth]{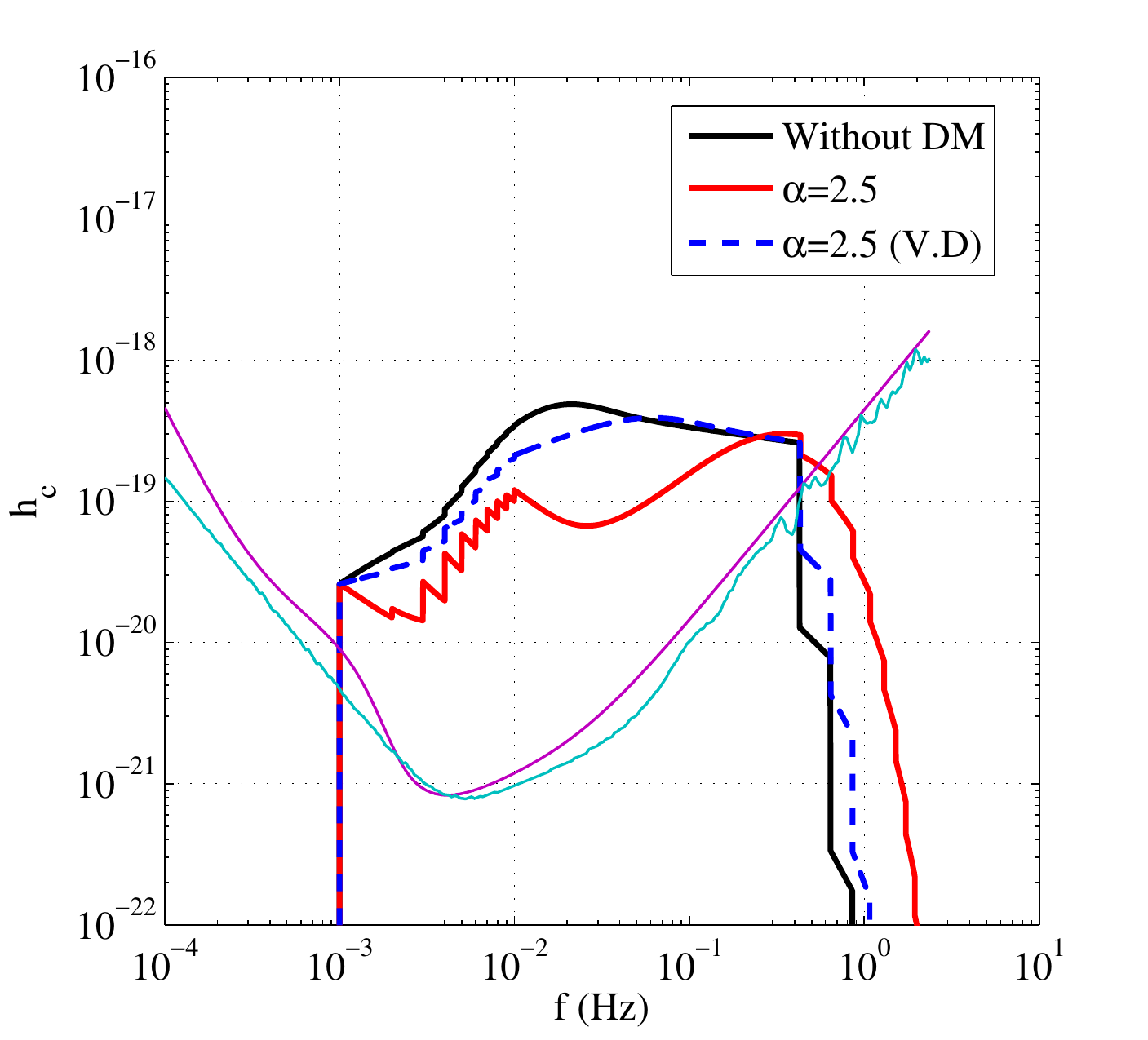}
	\end{minipage}
	\begin{minipage}[t]{\columnwidth}
		\includegraphics[width=\textwidth,height=0.95\textwidth]{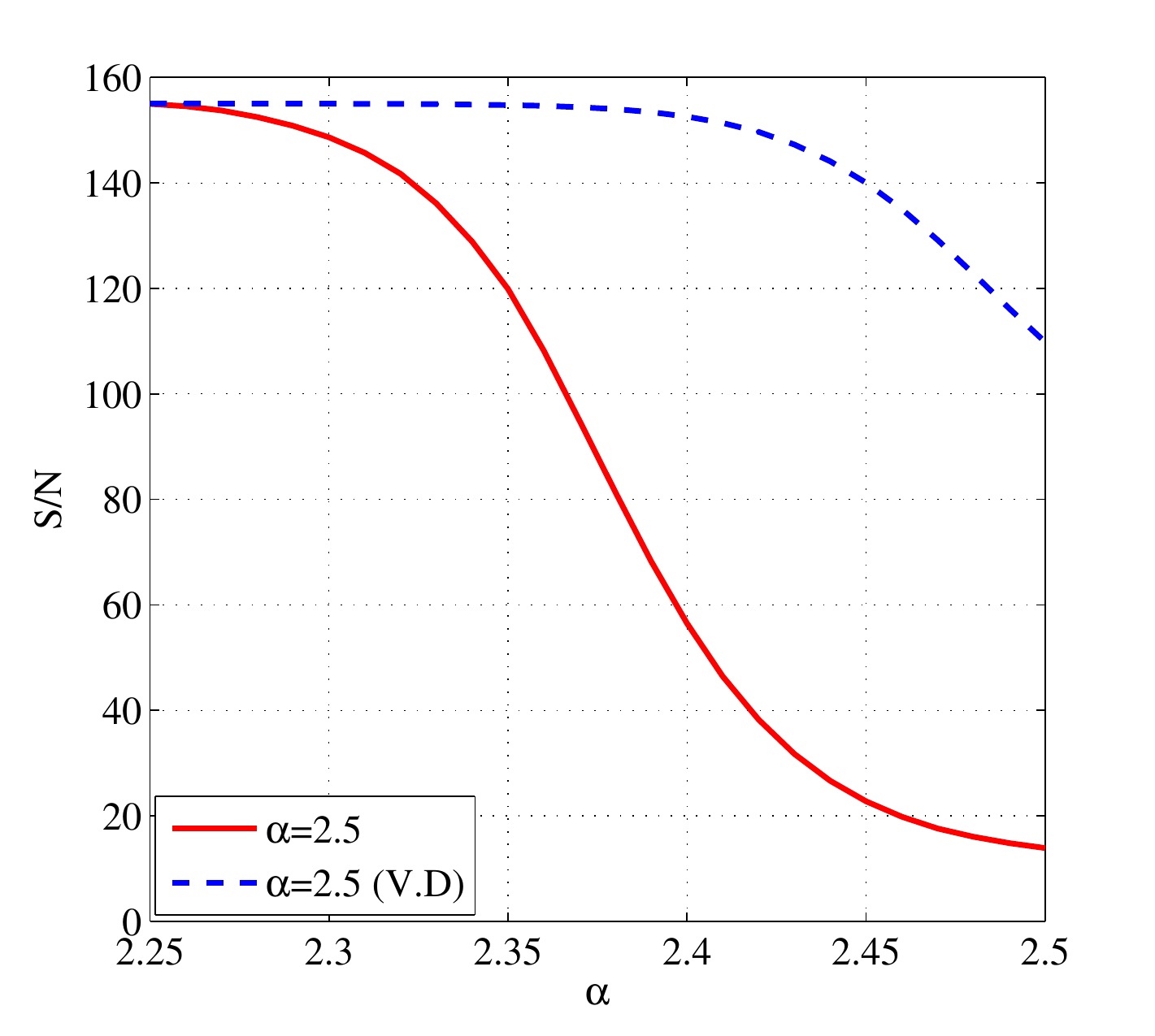}
	\end{minipage}  
	\caption{\label{fig:h_t_VD}(Left) GW characteristic strains when considering the effect of velocity dispersion. All the parameters are identical to those in Fig.~\ref{fig:harmonics}. The dashed blue curve shows the case with DM particles having velocity dispersion. (Right) $S/N$ plotted against the DM density index $\alpha$. Both curves are from the binary with parameters $m_1=10^{4}M_\odot$, $m_2=10 M_\odot$, $f_{\textrm{ini}}=10^{-3}/(1+z)$~Hz, $e_0=0.75$, and z=0.01. The dashed blue curve is obtained with velocity dispersion.}
\end{figure*}

In this section, we discuss the effects of DM particles with velocity dispersion on GW spectra and the signal-to-noise ratio. Note that the exact form of velocity dispersion is uncertain because of the lack of direct observational evidence of the vicinity of black holes. Here, we follow the standard treatment via the Eddington inversion procedure~\cite{binney2011galactic}. Using the phase space to describe the DM particle distribution, we can obtain the DM density distribution. Function~\cite{Kavanagh:2020cfn} is
\begin{eqnarray}\label{eq:VD}
	\rho_{_\textrm{DM}}(r) = \int_{0}^{v_{_\textrm{orb}}} v^2 f (E_r) d v,
\end{eqnarray}
where $E_r$ is the relative energy per unit mass, and $f(E_r)$ is the phase space distribution function. The upper limit of the integration is determined by the orbital velocity of the secondary inspiral object $v_{orb}$. Here, the integration indicates that only particles moving slower than the orbital velocity $v_{orb}$ contribute to dynamic friction. Note that $v_{orb}$ is determined by the orbital dynamics and is time-changing, as the secondary object moves along the elliptic orbit with typical velocity $v^2_{orb}\lesssim c^2/3$, so do the DM particles.

In Fig.~\ref{fig:h_t_VD} we show that the effects on GW characteristic strains and the signal-to-noise ratio from EMRIs can be modified by DM and velocity dispersion. In the left panel, we plot with all other parameters identical to those in Fig.~\ref{fig:harmonics}. The black curve represents the characteristic spectra from the GW of EMRIs or IMRIs without DM; the solid red curve is obtained with DM and neglecting velocity dispersion, and the dashed red curve is obtained by considering velocity dispersion. One can easily identify that the features in Section~\ref{sec:GW} remain, although the magnitude is now smaller. The reason is that the eccentricity, in this case, decreases faster than in the case without velocity dispersion, but it still decreases slower than in the case without DM. The evolution of eccentricity agrees with that in~\cite{Becker:2021ivq}.

We also show in the right panel of Fig. \ref{fig:h_t_VD} how the effect on the S/N curves of GWs changes when velocity dispersion is considered. The solid red curve is identical to that in Fig.~\ref{fig:SNR}, while the dashed blue one shows the modifications when velocity dispersion is considered. As expected, the magnitude of the modification is reduced but still sizable and relevant for future searches when the power index of the DM density profile $\alpha$ is larger than $\sim 2.4$.

\section{CONCLUSION}\label{sec:concl}

We have discussed the effects of DM on the binary system of a stellar BH or neutron star orbiting a massive BH surrounded by a DM spike. We have considered the cases with general elliptical orbits and demonstrated that a DM spike can substantially modify the evolution of the binary system and correspondingly affect GW emission.
In particular, we have found that the DF from DM can modify the GW's characteristic strain and frequency distribution, namely, suppression at low frequency but enhancement at a higher range. These effects will affect the detection and parameter inference of EMRI sources by reducing the signal-to-noise ratio at space-borne GW detectors, such as TAIJI and LISA. Extending the sensitivity to high frequencies near 1.0~Hz will be useful for future space-borne GW detectors. For various EMRIs, the impact of these effects can differ by up to one order of magnitude, depending on the DM density distribution, and is generally more substantial for smaller central black holes. Furthermore, we have investigated the possible reduction of these effects when considering DM's velocity dispersion. All of these results, which are illustrated in Figs.~\ref{fig:harmonics}, \ref{fig:SNR}, and \ref{fig:h_t_VD}, suggest that gravitational waves from EMRIs may provide a new probe for DM detection~\footnote{While we were finalizing the manuscript, two preprints~\cite{Becker:2021ivq, Dai:2021olt} appeared. Both papers discussed the effects of DM spikes on the evolution of eccentricity, which partly overlap our results in the middle of the right panel in Fig.~\ref{fig:harmonics}}.

\begin{acknowledgments}
	YT is supported by the National Key Research and Development Program of China (Grant No.2021YFC2201901), Natural Science Foundation of China (NSFC) under Grants No.~11851302, the Fundamental Research Funds for the Central Universities and Key Research Program of the Chinese Academy of Sciences, Grant No. XDPB15. YLW is supported in part by the National Key Research and Development Program of China under Grant No.~2020YFC2201501, and NSFC under Grants No.~11851302, No.~11851303, No.~11690022, No.~11747601, the Strategic Priority Research Program of the Chinese Academy of Sciences under Grant No. XDB23030100, and NSFC special fund for theoretical physics under Grant No. 12147103.
\end{acknowledgments}

\section*{Appendix}
We consider an EMRI system comprising a massive BH with a mass of $m_1$ and a compact object with a mass of $m_2$, $m_1\gg m_2$. We define the total mass as $M=m_1+m_2\simeq m_1$ and the reduced mass as $\mu =\frac{m_1m_2}{m_1+m_2}\simeq m_2$. Using Kepler’s law of an elliptic binary system, we have the relation for radius $r$, eccentricity $e$, and semilatus rectum $p$,
\begin{equation}
	r=\frac{p}{1+e\cos \varphi},\; e^2=1+\frac{2EL^2}{G^2M^2\mu^3},\; p=\frac{L^2}{GM\mu^2},
\end{equation}
where $E$ and $L$ are the energy and angular momentum of the system, respectively. Because of the GW emission and DF, $E$ and $L$ are evolving according to
\begin{equation}
	\frac{dE}{dt}= \langle \frac {dE}{dt }\rangle _{_\textrm{GW}} +  \langle \frac {dE}{dt} \rangle _{_\textrm{DF}},\;
	\frac{dL}{dt}= \langle \frac {dL}{dt }\rangle _{_\textrm{GW}}  +  \langle \frac {dL}{dt} \rangle _{_\textrm{DF}},
\end{equation}
where individual contributions are given by
\begin{widetext}
	\begin{equation}\label{eq:en}
		\begin{split}
			\langle \frac {dE}{dt }\rangle _{_\textrm{GW}} =&-\frac {32}{5} \frac {G^4\mu^2M^3} {c^5p^5} (1-e^2)^{3/2}(1+\frac{73}{24}e^2+\frac{37}{96}e^4)(1-e^2)^{3/2},\\
			\langle \frac {dE}{dt }\rangle _{_\textrm{DF}} =&	\frac{2G^{3/2} \mu^2\rho_{sp}r_{sp}^2 \ln \Lambda }{p^{(\alpha+1)}M^{1/2}} (1-e^2)^{3/2}\int_{0}^{2 \pi} d\varphi\frac{(1+e \cos \varphi )^{\alpha-2}} {{(1+2e\cos \varphi +e^2 )}^{1/2}}{\left[p-2R_s(1+e\cos{\varphi})\right]}^3,
		\end{split}
	\end{equation}
	and
	\begin{equation}
		\begin{split}\label{eq:l}
			\langle \frac {dL}{dt }\rangle _{_\textrm{GW}}=&-\frac {32}{5} \frac{G^{7/2}\mu^2{M}^{5/2}}{c^5p^{7/2}}  (1-e^2)^{3/2}(1+\frac{7}{8}e^2),\\
			\langle \frac {dL}{dt }\rangle _{_\textrm{DF}}=&\frac{2G \mu^2\rho_{sp}{r_{sp}}^\alpha \ln \Lambda (1-e^2)^{3/2} }{p^{\alpha+5/2} {M}} \int_{0}^{2 \pi} d\phi
			\frac{ (1+e\cos \varphi)^{\alpha-2} }{(1+2e\cos \varphi +e^2 )^{3/2}} (1+e \cos {\varphi} )^{\alpha-2}{\left[p-{2R_s}{(1+e\cos\varphi})\right]^3} ,
		\end{split}
	\end{equation}
	where $\ln \Lambda \approx 10$ is the Coulomb logarithm~\cite{Amaro_Seoane_2007} due to DF. Using the relations $p=a(1-e^2)$, $a = - Gm\mu/(2E)$, and $f_{\textrm{orb}}=\sqrt{GM/a^3}/(2\pi)$, we can obtain the differential equations for orbital frequency and eccentricity,
	\begin{equation}\label{eq:f}
		\begin{split}
			\frac{df_{\textrm{orb}}} {dt} =& \frac{96}{5}\frac{G^{5/3}{(2\pi )}^{8/3}\mu{M}^{2/3}f_{\textrm{orb}}^{11/3}}{c^5}{(1 - {e^2})^{ - 7/2}}(1 + \frac{{73}}{{24}}{e^2} + \frac{{37}}{{96}}{e^4}) + \frac{{{6(2\pi )}^{2/3\alpha  - 1}G^ {1- \alpha /3}u{\rho _{sp}}{r_{sp}}^\alpha \ln \Lambda }}{{{{M}}^{ 1 + \alpha/3 }}}{f_{\textrm{orb}}^{2\alpha /3}}\\
			&{(1 - {e^2})^{ - \alpha  - 1}}
			\int_0^{2\pi } {d\varphi } \frac{{{{(1 + e\cos \varphi )}^{\alpha  - 2}}}}{{{{({e^2} + 2e\cos \varphi  + 1)}^{1/2}}}}{\left[(1 - {e^2}) - \frac{{4{{(2\pi )}^{2/3}}{{(G{M})}^{2/3}}{f_{\textrm{orb}}^{2/3}}}}{{{c^2}}}(1 + e\cos \varphi )\right]^3},
		\end{split}
	\end{equation}
	\begin{equation}\label{eq:e}
		\begin{split}
			\frac{de}{dt} = & -\frac{{304}}{{15}} \frac{G^{8/3} \mu {M^{4/3}} {(2\pi )}^{8/3} {f_{\textrm{orb}}}^{8/3} } {{c^5}} e{(1 - {e^2})^{ - 5/2}} (1 + \frac{{121}}{{304}}{e^2}) - \frac{4{(2\pi )}^{2\alpha /3 - 1} G^{ 1- \alpha /3} u{\rho _{sp}}{r_{sp}}^\alpha \ln \Lambda } {M^{\alpha /3}}{{f_{\textrm{orb}}}^2}{^{\alpha /3 - 1}}\\
			& (1 - {e^2})^{ - \alpha}\int_0^{2\pi } {d\varphi } \frac{(e + \cos \varphi )  {(1 + e\cos \varphi )}^{\alpha  - 2}    }  {{({e^2} + 2e\cos \varphi  + 1)}^{3/2}}\left[(1 - {e^2}) - \frac{ 4{(2\pi )}^{2/3}{(GM)}^{2/3}{f_{\textrm{orb}}}^{2/3}   } {{c^2}} (1 + e\cos \varphi )\right]^3.
		\end{split}
	\end{equation}
	Notably, the relation
	\begin{equation}\label{eq:eEL}
		e\frac{de}{dt}=\frac{p}{Gm\mu }\frac{dE}{dt}+\frac{e^{2}-1}{\sqrt{Gm\mu^{2}p}}\frac{dL}{dt},
	\end{equation}	
	ensures $\frac{de}{dt}\leq 0$ in the absence of DF. One can easily check $\langle \frac {dE}{dt }\rangle _{_\textrm{GW}}< 0, \langle \frac {dL}{dt }\rangle _{_\textrm{GW}}<0$ and their relative size. However, because of the factor $e^2-1<0$ in the front of the second term above, $\frac{de}{dt}$ can be positive in the presence of dynamical friction when the absolute value of $\langle \frac {dL}{dt }\rangle _{_\textrm{DF}}$ is large.
	
	The function $g_n(e)$~\cite{Peters_1963} is defined by
	\begin{eqnarray}\label{eq:gn}
		g_n(e) = n^4&\Bigg\{\Big[ J_{n-2}(ne)-2eJ_{n-1}(ne)+\frac{2}{n}J_n(ne)
		+2eJ_{n+1}(ne) -J_{n+2}(ne)\Big]^2  \\
		&+(1-e^2)\Big[J_{n-2}(ne)-2J_n(ne)+J_{n+2}(ne)\Big]^2 +\frac {4}{3n^2}{J_n}^2(ne)\Bigg\}.
	\end{eqnarray}
\end{widetext}



%

\end{document}